\documentclass[letterpaper]{VisionStyle}
\usepackage{epsfig}

\begin{document}

\title{X-ray Line Emission in Hercules X-1}

\author{M.A.\,Jimenez-Garate\inst{1} \and C.J.\,Hailey\inst{2} \and J.W.\,den Herder\inst{3}
\and S.\,Zane\inst{4} \and G.\,Ramsay\inst{4}}

\institute{
MIT Center for Space Research, 77 Vassar St. Room 37-274, NE80-6091, Cambridge, MA, 02139, USA
\and 
Columbia Astrophysics Laboratory, 550 West 120th St, New York, NY, 10027, USA
\and
Space Research Organization of The Netherlands, Sorbonnelaan 2, 3584 CA Utrecht, The Netherlands
\and
Mullard Space Science Laboratory, University College London, Holmbury St. Mary, Dorking, Surrey, RH5 6NT, England, UK
}

\maketitle 

\begin{abstract}
We find line emission from the hydrogen- and/or helium-like ions of Ne, O, N and
C in the low and short-on states of Her X-1, using the \it XMM-Newton
\rm Reflection Grating Spectrometer.
The emission line velocity broadening is $200 < \sigma < 500$~km~s$^{-1}$.
Plasma diagnostics with the
\ion{Ne}{IX}, \ion{O}{VII} and \ion{N}{VI} He$\alpha$ lines and
the radiative recombination continua of \ion{O}{VII} and \ion{N}{VII},
indicate the gas is heated by photoionization, with an
electron temperature of $T = 6 \pm 2$~eV.
We use spectral models to measure element abundance ratios of
$[{\rm N}/{\rm O}] = 9.5 \pm 1.3$, $[{\rm C}/{\rm O}] = 0.68 \pm 0.16$,
and $[{\rm Ne}/{\rm O}] = 2.4 \pm 1.2$
times solar, which quantify CNO processing in HZ Her.  
Photoexcitation and high-density effects are not differentiated by the  
measured He$\alpha$ lines, as shown by calculations that use previous UV
photometry.  We set limits on the location of
the line emission region, and find an electron density $n_e >
10^{11}$~cm$^{-3}$.  The narrow emission
lines can be attributed to reprocessing in either an accretion disk
atmosphere and corona or on the X-ray illuminated face of HZ Her. 
In the main-on state, the bright continuum only allows the detection
of interstellar absorption, plus \ion{O}{VII} He$\alpha$ emission lines 
with $\sigma = 3200 \pm 700$~km~s$^{-1}$
and complex profiles. Other broad lines may be present.
The broad lines may originate in a region near the pulsar magnetosphere.
In such a case, periodic occultations of the Her X-1 magnetosphere
must occur with 35~d phase, consistent with the precession of the accretion disk.
Fe L lines are not detected.

\end{abstract}

\section{Introduction}

In Her X-1, X-rays are being generated by the infall
of plasma from a $2.3 ~M_{\sun}$ star (HZ Her) onto a $1.5 \pm 0.3 ~M_{\sun}$ neutron star
(\cite{mjimenez-C1:rey97}).
The neutron star pulsates in
the X-ray band with period $P_{\rm pulse} =1 \fs 24$ (\cite{mjimenez-C1:tan72}).
The thermal emission observed in the soft X-ray band, 
the large luminosity of the system ($L = 3.8 \times 10^{37}$ erg s$^{-1}$),
and the observed broad UV emission lines (\cite{mjimenez-C1:bor97}),
all indicate the mass transfer is being mediated by an accretion disk. 
Her X-1 is unusually well placed for soft X-ray and UV observations,
since it is $\sim 3$~kpc above the galactic plane while at a distance
of $6.6 \pm 0.4$~kpc (\cite{mjimenez-C1:rey97}).
An orbital period of $P_{\rm orb} = 40 \fh 8$ can be measured from
X-ray eclipses (\cite{mjimenez-C1:tan72}).

One of our main goals is to find the spectral signatures
of a precessing accretion disk.
The Her X-1 system goes through high and low X-ray flux states in an almost
periodic fashion, with a $P_{\psi} = 35$~d pseudo-period
(\cite{mjimenez-C1:gia73}).
Variability as a function of $\psi$ has also been observed in the optical
light curves (\cite{mjimenez-C1:ger76}), X-ray pulse shapes (\cite{mjimenez-C1:dee91}),
X-ray dips (\cite{mjimenez-C1:sco99}), and X-ray spectra (\cite{mjimenez-C1:ram02}
and this paper). The $\psi$-phase has been associated with
disk precession (\cite{mjimenez-C1:pet91,mjimenez-C1:sco00}).
Only a handful of other X-ray sources are known to exhibit such
pseudo-periodicities, and Her X-1 is the brightest of the group.
The binary system is being observed nearly edge-on, at an inclination angle
of $i \sim 85^\circ$. Models of the disk atmosphere and corona (\cite{mjimenez-C1:jim01}) show
that its line emission is detectable, especially at large $i$ and
in binaries with large ($\sim 10^{11}$~cm) disks.
Her X-1 is also a prime object to study the interaction
of an accretion disk with a strong magnetic field ($B = 3.5 \times 10^{12}$ G from
\cite{mjimenez-C1:dal98}, \cite{mjimenez-C1:tru78}).

\section{Spectroscopy with RGS}

Three observations with the \it XMM-Newton \rm Reflection
Grating Spectrometer ($RGS$), during the low, short-on and main-on states,
reveal recombination line and continuum emission in Her X-1.
We find dramatic spectral changes through the 35~d cycle, including changes
in the continuum, line fluxes, and line widths.  The high resolution
X-ray spectra unveil two new components
inside the Her X-1 system:
\begin{enumerate}
\item A photoionized narrow-line region, with velocity broadening
$\sim 300$~km~s$^{-1}$.  The density of this material is
$n_{\rm e} > 10^{11}$~cm, with the lowest
temperatures in the $20,000 < T < 70,000$~K
range. We set an upper bound for the radius of this region of
$r < 10^{12}$~cm, just outside the $3 \times 10^{11}$~cm Roche
lobe radius.
\item A photoionized broad-line region evidenced by \ion{O}{VII} He$\alpha$ lines
with velocity broadening $\sim 3,500$~km~s$^{-1}$.
If the broadening is due to orbital velocity,
this region is between $5 \times 10^8 < r < 3 \times 10^9$~cm.
We estimate the density of this region to be 
$10^{16} < n_{\rm e} <  10^{19}$~cm$^{-3}$.
\end{enumerate}
\subsection{Low and short-on states: narrow line emission}

The low and short-on state spectra consist of a power-law continuum
plus radiative recombination ($RR$) emission.
The brightest line features have a velocity broadening
$200 < \sigma < 500$~km~s$^{-1}$.
The count rate in the short-on is twice 
as that in the low state, for both continuum and most lines. The \ion{N}{VI},
\ion{O}{VII}, and \ion{Ne}{IX} triplet ratios indicate the gas
is photoionized (section \ref{mjimenez-C1:secdiag}).  The power-law continuum is
probably due to Compton scattering of the neutron star
X-rays in the accretion disk corona and/or
on the face of the companion. 

The variation of the X-ray line
flux with 35~d phase, vis-a-vis the expected
accretion disk inclination from \cite*{mjimenez-C1:sco00},
suggests that at least half of the line emission originates on
the face of the companion. If the accretion disk atmosphere and
corona are the only source of X-ray lines, then we expect the
outer-disk inclination $i_{\rm d}$ to be in disagreement
with the \cite*{mjimenez-C1:sco00} values.
As the disk precesses from $i_{\rm d} \sim 90^\circ$ to
$i_{\rm d} \sim 80^\circ$, the
observed lines should increase their flux and broadening, because
the inner-disk becomes unobstructed
by the outer-disk (\cite{mjimenez-C1:jim01}).
According to the $i_{\rm d} (\psi)$ from \cite*{mjimenez-C1:sco00},
the disk was more edge-on during the short-on than during
the low state observation. If the line emission originates
on the HZ Her, this can be explained by shadowing from the disk.
\begin{figure}[ht]
  \begin{center}
    \epsfig{file=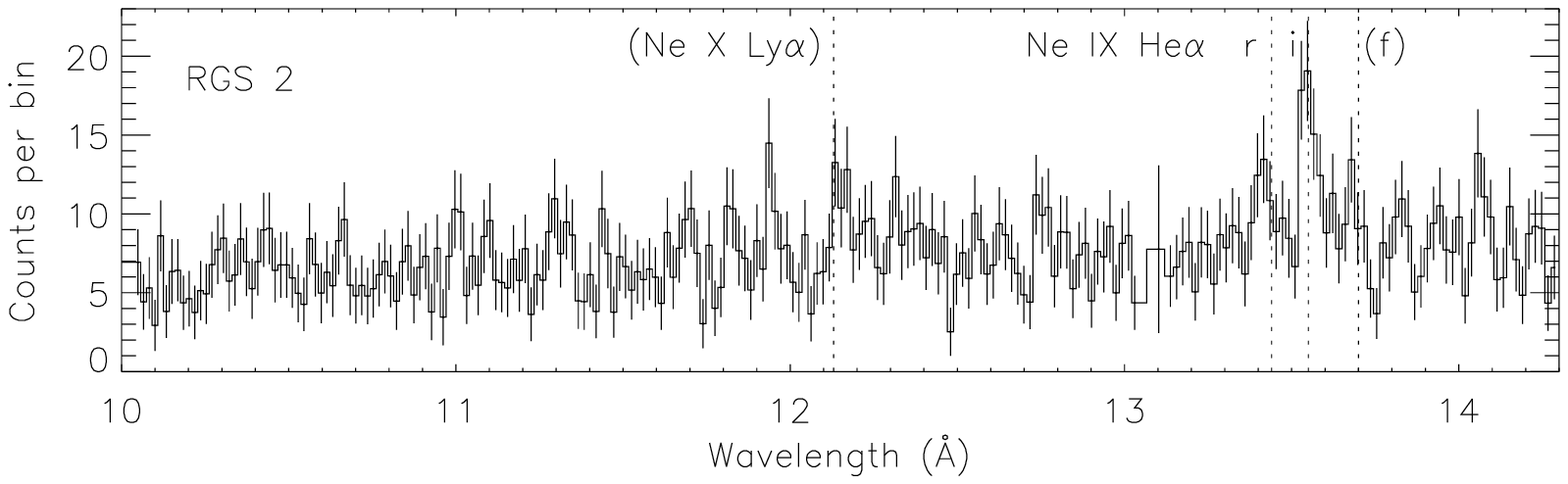, width=8.5cm}
    \epsfig{file=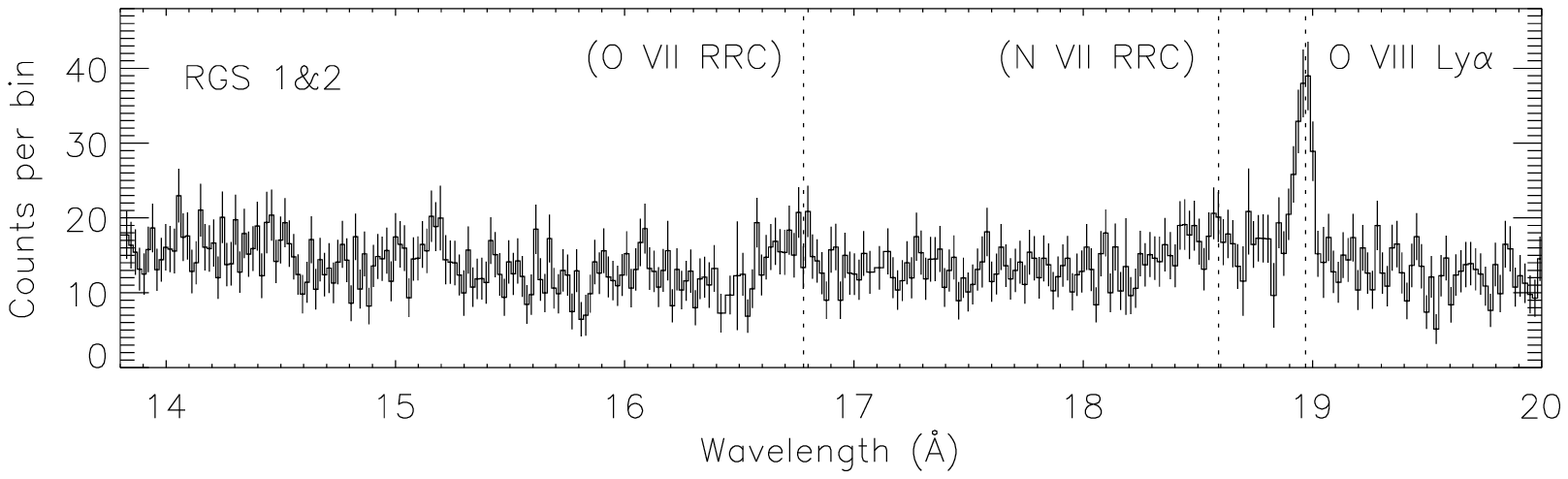, width=8.5cm}
    \epsfig{file=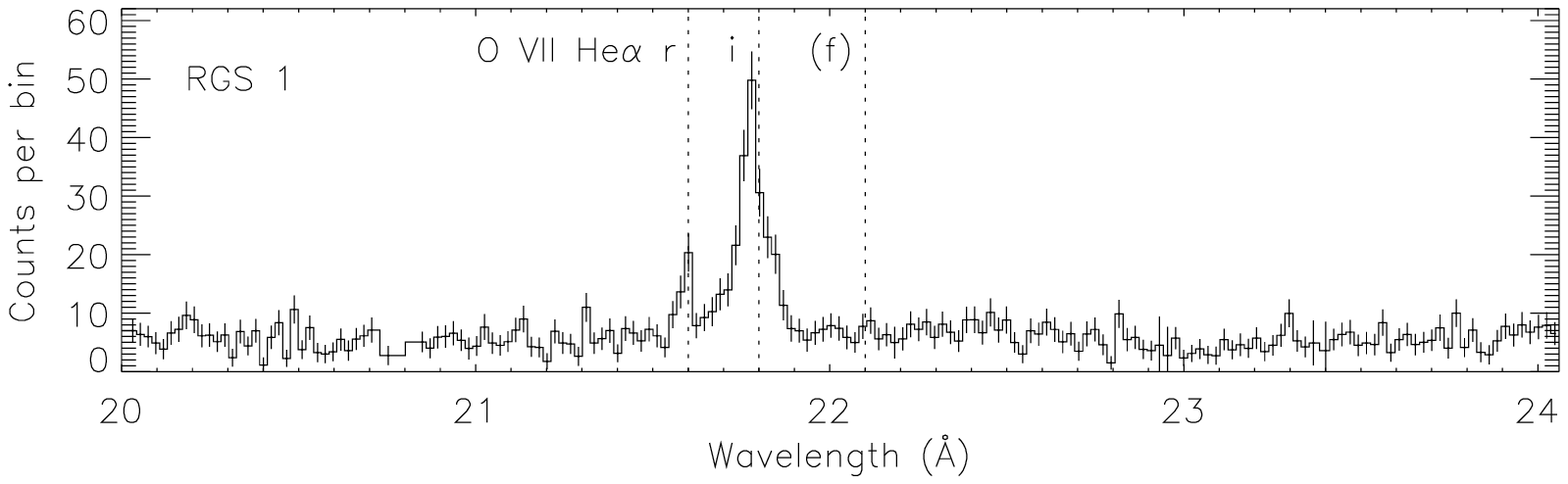, width=8.5cm}
    \epsfig{file=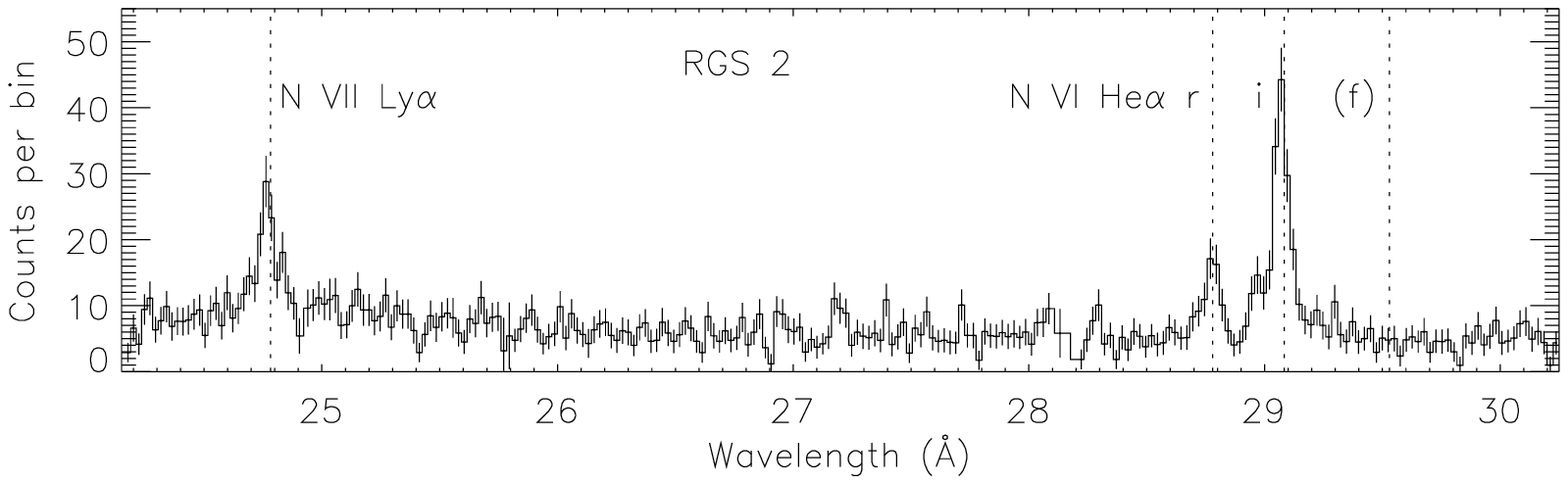, width=8.5cm}
    \epsfig{file=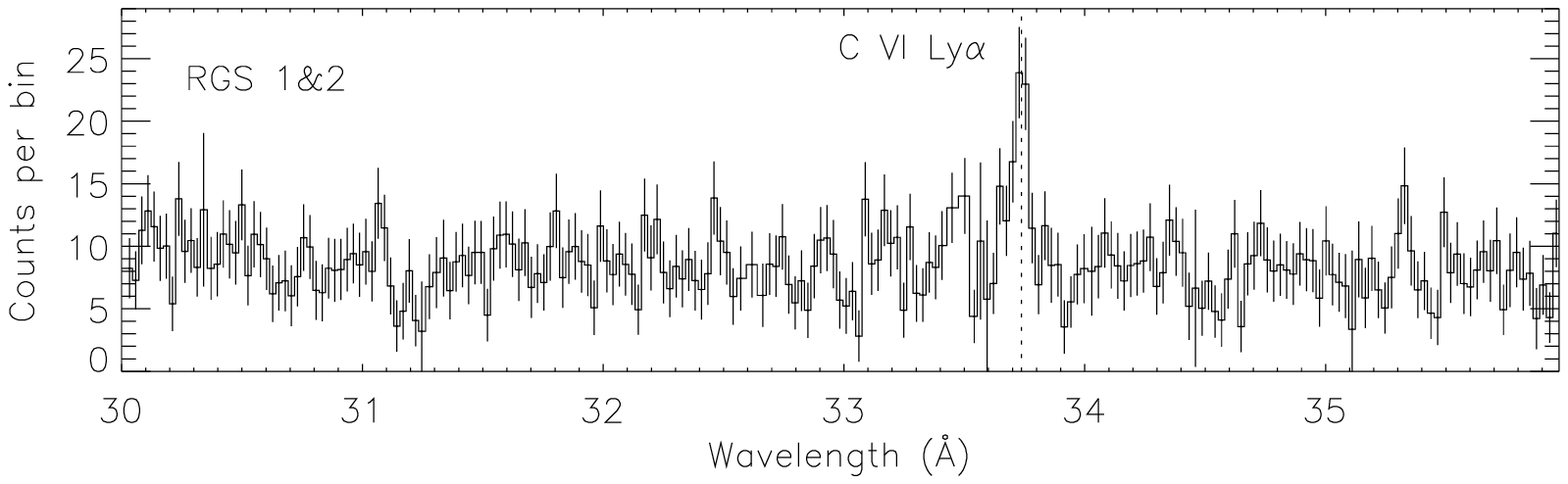, width=8.5cm}
  \end{center}
\caption{The added $RGS$ spectrum of low and short-on state observations (of 12 and 13 ks).
The bin size is 18, 21, 23, 25, and 27~m\AA \ for each successive
frame, from short to long wavelengths. 
 }
\label{mjimenez-C1_fig:lowsh}
\end{figure}

\subsection{The main-on: magnetosphere emission?}
\begin{figure}[ht]
  \begin{center}
    \epsfig{file=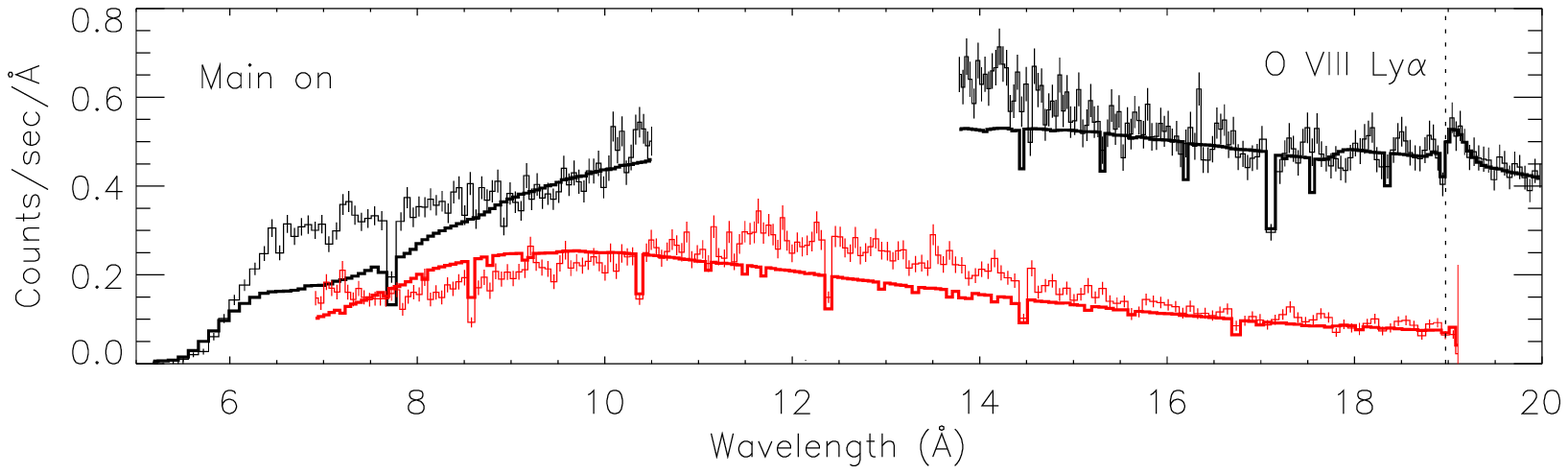, width=8.8cm}
    \epsfig{file=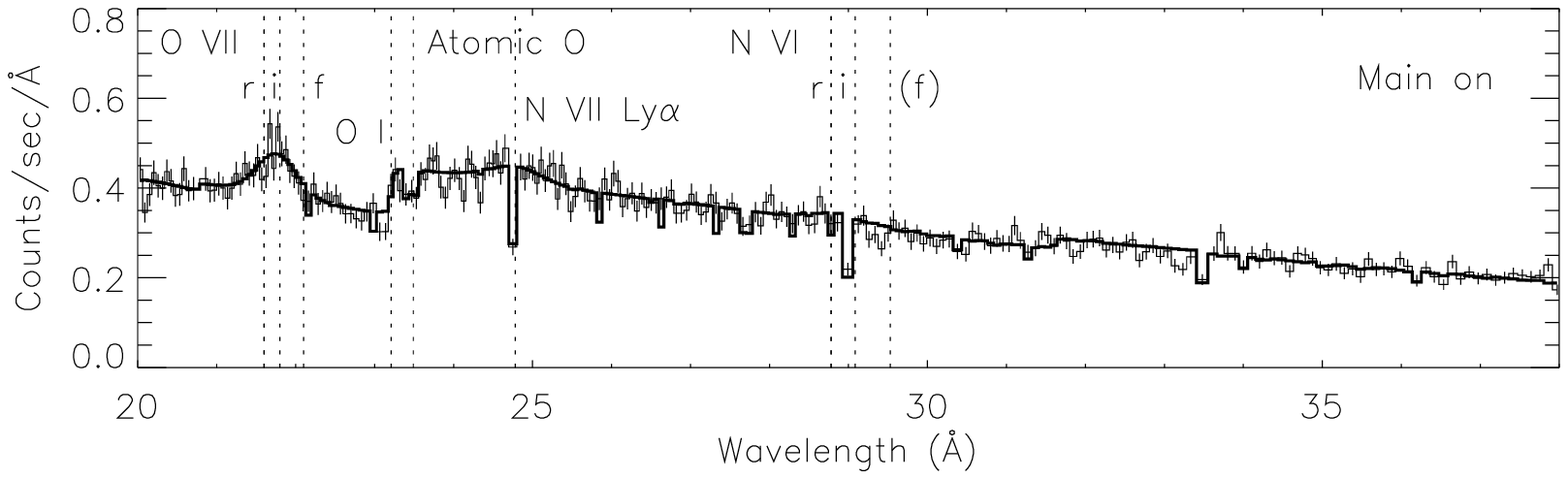, width=8.8cm}
  \end{center}
\caption{ High-resolution spectrum with $RGS$ 1 (histograms with error bars)
in the main-on state observation at orbit 207 (11~ks). The power-law
continuum model (thick-line histograms) is meant to emphasize the
"bump" at 10--15~\AA. Both first (black) and second order (red) spectra
are shown. }
\label{mjimenez-C1_fig:main}
\end{figure}

The main-on state in the 5--38~\AA \ band
is dominated by a smooth continuum and a broad continuum
"bump" previously identified with Fe L line emission
(see figure \ref{mjimenez-C1_fig:main}). Most of the soft X-ray continuum
originates from reprocessing on the disk, judging
by its sinusoidal pulse profile. 
The sharply peaked pulse profile observed below 
$\sim 10$~\AA \ (\cite{mjimenez-C1:ram02}), indicates
the emission region is compact ($< 10^8$~cm), and near the neutron star.
The 10--15~\AA \ "bump" does not contain any
obvious Fe L lines. The continuum shape is complex and
has not been successfully fitted yet with a model.

\begin{figure}[ht]
  \begin{center}
    \epsfig{file=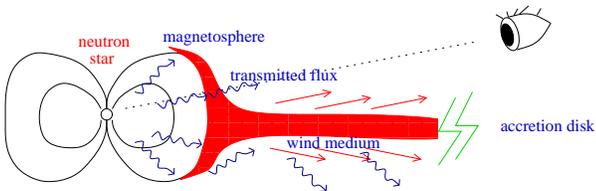, width=8cm}
  \end{center}
\caption{ Sketch of the hypothetical inner region of the accretion disk and its
interface with the magnetosphere. 
The broad line region coincides with the magnetosphere or the
neighboring disk.
The 1~keV "bump" could be an absorption
complex from the photoionized gas at the base of the magnetosphere,
which can be observed at high inclination.
P Cygni profiles may form in the disk wind for optically thick
lines (i.e. \ion{O}{VIII} Ly$\alpha$).  The disk is warped and its scale height
increases with radius (not shown).  The scale is $\sim 10^9$~cm
(neutron star radius $\sim 10^6$~cm). }
\label{mjimenez-C1_fig:mag}
\end{figure}

Weaker, discrete spectral features are present as well.
A surprisingly broad \ion{O}{VII} He$\alpha$ line complex, with
$\sigma = 3200^{+600}_{-700}$~kms$^{-1}$, 
is centered on the $i$ line, and is consistent
with the line ratios observed in the low and short-on states.
A weaker, similarly broad \ion{N}{VII} Ly$\alpha$ line may also be present.
High density plasma funneled out of the disk by the magnetic field
forms a magnetospheric shell which 
may produce recombination emission as it
is blasted by pulsar X-rays.
If the magnetospheric ions are fully stripped of electrons,
photons which are Compton scattered by those electrons
may illuminate the neighboring disk and produce
recombination (see figure \ref{mjimenez-C1_fig:mag}).
A 1s-2p resonant absorption feature from
atomic O is detected.
A neutral O edge is mostly instrumental
in nature. The second order effective area also seems overestimated
by the calibration below $\sim 10$~\AA,
if we take the first order as reference.
There is marginal evidence for a complex
line shape, possibly a P Cygni with $v>2000$~km~s$^{-1}$,
in \ion{O}{VIII} Ly$\alpha$ (figure \ref{mjimenez-C1_fig:main}).
P Cygni profiles for winds with velocities of $v\sim 800$~km~s$^{-1}$ 
have been found by \cite*{mjimenez-C1:chi00} in the UV.
The upper bounds for the fluxes of unresolved lines are below those observed in
the low and short-on states, for many of the lines
(see figure \ref{mjimenez-C1_fig:lowsh}). This
may be due to an orbital phase ($\phi$) dependence of the line fluxes,
since the main-on observation was performed at $\phi \simeq 0.22$,
while the other two observations were obtained at $\phi \simeq 0.50$,
suggesting the narrow lines originate on HZ Her.

\section{Plasma Diagnostics}
\label{mjimenez-C1:secdiag}

We use five kinds of spectral diagnostics:
\begin{enumerate}
\item The $G$ ratio in He-like ion line triplets.
The triplet consists of a resonance ($r$), intercombination ($i$) and
a forbidden ($f$) line.  The plasma is found to be dominated by
photoionization since the $G=(i+f)/r \simeq 4.2$. In contrast,
collisional plasmas have $G \sim 1$ (\cite{mjimenez-C1:lie99}).
\item The $R=f/i$ line ratio as density diagnostic in He-like ions. 
When the density is higher than some threshold, $R \to 0$,
while in the low density limit, $R \sim 4$ 
(\cite{mjimenez-C1:por00}). We observe $R < 0.1$.
\item The $R$ ratio as UV radiation probe in He-like ions. 
Excitation by UV photons can also explain $R \to 0$
(\cite{mjimenez-C1:mew78}, \cite{mjimenez-C1:kah01}).
We find photoexcitation competes with electron
impact excitation. Using UV photometry
from \cite*{mjimenez-C1:bor97}, we
set $r < 10^{12}$~cm for the radius of the line emission region.
\item Temperature diagnostics with the radiative recombination
 continuum ($RRC$) (\cite{mjimenez-C1:lie96}). The width of the $RRC$
is proportional to the electron temperature. For \ion{O}{VII}
and \ion{N}{VII}, we find $kT=4 \pm 2$~eV and $kT = 6 \pm 2$~eV, respectively. 
Model calculations with $XSTAR$ v.2.1d (\cite{mjimenez-C1:kal82})
show $2 < T < 4$~eV for \ion{N}{VII} and \ion{O}{VII}. 
\item Using the ionization parameter, and assuming thermal and
ionization balance in the optically thin case, we constrain the
density to $n_{\rm e} > 4 \times 10^{11}$~cm$^{-3}$ for the narrow line region,
and $4 \times 10^{16} < n_{\rm e} < 10^{19}$~cm$^{-3}$ for the broad
line region of the main-on state.  We use the limits on the distance
to the pulsar that we set from photoexcitation and from velocity broadening.
\end{enumerate}
\section{Emission measure analysis}
  
This analysis allows us to measure or set limits on the density,
geometry, and element abundances of the emission regions.
We use plasma and spectral models together with a phenomenological 
emission measure distribution to fit the observed line
fluxes.  The emission measure, $EM = \int n_{\rm e}^2 dV$, equals
the volume times the density squared. The $EM$ is proportional
to the recombination line flux. The $EM$ may vary
as a function of the ionization parameter $\xi = L/n_er^2$. 
To fit the line fluxes, we need
the differential emission measure $DEM = d(EM)/d(\log_{10} \xi)$
(as defined in \cite{mjimenez-C1:lie99}).
We use recombination rates calculated by Liedahl (private communication)
with the $HULLAC$ atomic code (\cite{mjimenez-C1:kla77}).

\begin{figure}[ht]
  \begin{center}
    \epsfig{file=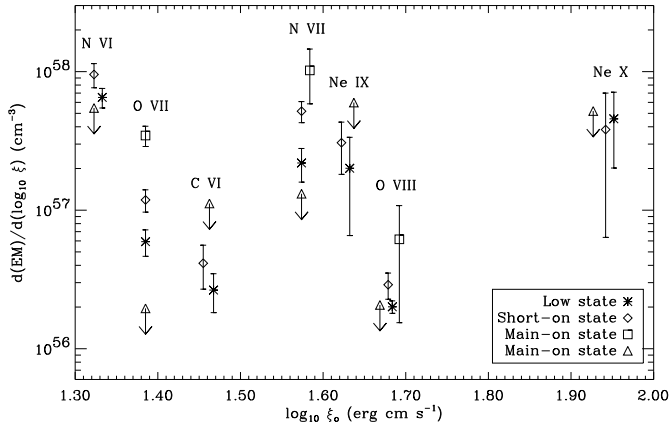, width=8.8cm}
  \end{center}
\caption{
The differential emission measure ($DEM$) for the line emission
detected at various Her X-1 states, versus the ionization
parameter which maximizes the line power. We create a 
model with a flat $DEM$ from $0.0 < \log_{10} \xi < 3.5$.
Upper limits for the $DEM$s of unresolved lines in the main-on are
denoted by triangles.
 }
\label{mjimenez-C1_fig:em}
\end{figure}

In one model, the emission measure is constant with respect to 
$\xi$, while in a second model, we
let the $DEM$ vary as a power law of $\xi$. 
The results of model 1, shown in figure \ref{mjimenez-C1_fig:em}, show 
that element abundance ratios differ from the
solar values. The $DEM$ of N and O is seen to depend on the 
charge state, invalidating model 1. 
The line flux variability, however, is well-represented by figure \ref{mjimenez-C1_fig:em}.
Unlike the flat-$DEM$ model 1, the power-law $DEM$ model 2 achieves
self-consistency, and from it, element abundance ratios are extracted.
We find $[{\rm C}/{\rm O}] = 0.68 \pm 0.16$,
$[{\rm N}/{\rm O}] = 9.9 \pm 1.3$ and $[{\rm Ne}/{\rm O}] = 2.4 \pm 1.0$ times solar for the low state
data, using the solar abundances from \cite*{mjimenez-C1:wil00}.
Nitrogen is significantly enriched,
indicating CNO processing in HZ Her.
A lower limit of $n_{\rm e} > 10^{14}$ cm$^{-3}$ is estimated from model 1
for the broad line region, by using the maximum volume deduced
from velocity broadening. 

\section{Conclusions}

The X-ray emission lines from Her X-1 are weak, but they provide a wealth
of information on gas which is too hot to observe at other wavelengths
and which may be very close to the neutron star.
By use of models, the X-ray spectrum yields more accurate element abundance ratios
between C, N, O, and Ne, constraining the historical nuclear burning in HZ Her.
The temperature can be measured, and limits on the density and 
geometry of the gas are set.  Two dynamically distinct photoionized
regions are identified. A low dispersion velocity region may be the accretion
disk atmosphere and corona, or the illuminated face of HZ Her. The high dispersion
velocity region, albeit weakly detected, may be dense material in the magnetosphere or
its neighboring inner-disk. 

The high resolution X-ray spectrum of Her X-1 varies dramatically with 35~d phase,
presumably due to the precession of the accretion disk.
This spectral variability may help us determine the
inclination-dependence of spectra from other X-ray binaries with disks.

Many features known to be present in the spectrum of photoionized gases have not been
detected due to the low signal-to-noise of these data. In particular, the Lyman series 
lines provide constraints on the column density and line optical depth, and
accurate measurements of the $RRC$ verify the validity of the plasma 
equilibrium calculations used here. This will allow us to identify the source of the
line emission and further investigate the environment around accreting neutron stars.

\end{document}